\documentclass[12pt,a4paper]{article}
\usepackage{cmap}
\usepackage[T2A]{fontenc}
\usepackage[utf8]{inputenc}
\usepackage{amsmath, amssymb, amsthm}
\usepackage{graphicx}
\usepackage{flafter}
\usepackage{geometry}
\graphicspath{{figs/}}
\title{PERCOLATION OF DISCORECTANGLES ORIENTED PREDOMINANTLY ALONG TWO MUTUALLY PERPENDICULAR DIRECTIONS}
\author{Yu.\,Yu.\,Tarasevich}
\date{Astrakhan Tatishchev State University\\ Astrakhan, Russia,
414056, 20a Tatishchev St.\\
Tel.:~(+7\,8512)\,246496 ext.~267, e-mail: tarasevich@asu-edu.ru
}

\begin{document}

\maketitle

In a recent work~\cite{Grazioli2026}, percolation of zero-thickness sticks oriented predominantly along two mutually perpendicular directions was studied. Both computer simulation and analytical estimates were used. The study revealed some universal properties of percolation, but the question remained open whether universality persists in the case of objects of nonzero thickness. The present study is intended to answer this question.

\begin{figure}[!htb]
  \centering
  \includegraphics[width=0.35\textwidth]{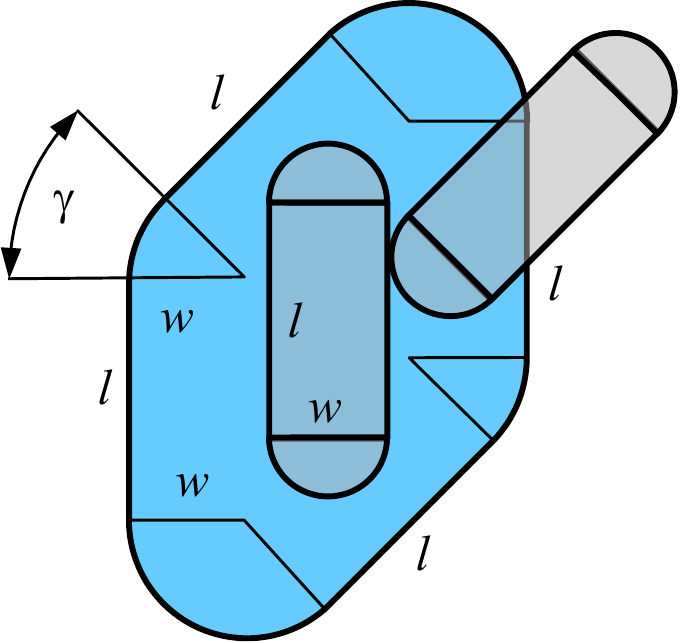}
  \caption{Discorectangle with dimensions indicated and the excluded area for two discorectangles with a given orientation}\label{fig:DR}
\end{figure}

A convenient elongated object of finite thickness for simulation is a discorectangle (stadium). A discorectangle is characterized by its total length $L$, width $w$ (diameter of the semicircles), and straight-section length $l$, with $L = l + w$ (Fig.~\ref{fig:DR}).

In dimensionless units, it is convenient to fix $L=1$ and specify the aspect ratio $\varepsilon = L/w$. Then
\[w = \frac{L}{\varepsilon}, \quad l = L - w = L\left(1 - \frac{1}{\varepsilon}\right).\]

The excluded area $A_{\text{ex}}$ for a pair of discorectangles with relative orientation angle $\gamma = \theta_j - \theta_i$ is calculated analytically~\cite{Balberg1984}:
\[A_{\text{ex}}(\gamma) = 4lw + \pi w^2 + l^2 |\sin \gamma|.\]

The authors of~\cite{Grazioli2026} used three families of distributions $\psi(\theta)$ satisfying the symmetry conditions $\langle S \rangle = 0$ (where $S = \cos 2\theta$) and $\psi(\theta) = \psi(\pi/2 - \theta)$. Since, as the authors showed, percolation properties are insensitive to the specific form of the distribution, we restrict ourselves to one angular distribution, which the authors of~\cite{Grazioli2026} call the “Maltese cross”. This distribution is specified by the parameter
\begin{equation}\label{eq:APDF-MC}
    \psi_{\text{mc}}(\theta, \alpha) =
    \begin{cases}
    \dfrac{1}{4\alpha}, & \theta \in [-\alpha, \alpha] \cup [\pi/2-\alpha, \pi/2+\alpha], \\
    0, & \text{otherwise},
    \end{cases}
     \qquad \alpha \in \left(0,\frac{\pi}{4}\right].
\end{equation}
For this distribution, the second moment of the orientational parameter is
\[\langle S^2 \rangle = \frac{1}{2}\left(1 + \frac{\sin(4\alpha)}{4\alpha}\right),\]
and the mean value of the modulus of the sine of the angle difference between two independent particles is calculated analytically
\[\langle |\sin \gamma| \rangle_{\text{mc}} = \frac{1}{4\alpha^2} \left( 2\alpha + 1 - \sin(2\alpha) - \cos(2\alpha) \right).\]
Limiting cases: as $\alpha \to 0$ we have $\langle |\sin \gamma| \rangle \to 1/2$; at $\alpha = \pi/4$ we have $\langle |\sin \gamma| \rangle=2/\pi$.

One of the analytical methods for estimating the percolation threshold used in~\cite{Grazioli2026} is the eigenvalue method. The angular interval $[-\pi/2, \pi/2)$ is divided into $N_\theta$ bins of width $\Delta\theta = \pi/N_\theta$. The bin weights $p_j$ are calculated as the integral of the density:
\[p_j = \int_{\theta_j - \Delta\theta/2}^{\theta_j + \Delta\theta/2} \psi(\theta) \, d\theta, \quad \sum_{j=1}^{N_\theta} p_j = 1.\]
A symmetric connectivity matrix $M$ is constructed:
\[M_{jk} = \sqrt{p_j p_k} \, A_{\text{ex}}(\theta_j - \theta_k).\]
In the tree-like network approximation, the percolation threshold is determined through the maximum eigenvalue of the connectivity matrix:
\[\rho_c = \frac{1}{\lambda_{\max}(\mathrm{M})}, \quad n_c = \rho_c L^2.\]

At $l=0$, $w=L$ (disks), the excluded area is independent of angle: $A_{\text{ex}} = \pi L^2$. The matrix $\mathrm{M}$ degenerates into a rank-1 matrix, and its only nonzero eigenvalue is $\lambda_{\max} = \pi L^2$. The dimensionless percolation threshold is
\[n_c = \frac{1}{\pi} \approx 0.3183.\]
At $w \to 0$, $l=L$ (zero-thickness stick), the excluded area is $A_{\text{ex}} = L^2 |\sin \gamma|$.
For the isotropic distribution ($\alpha = \pi/4$): $\lambda_{\max} = \frac{2}{\pi} L^2$, $n_c = \frac{\pi}{2} \approx 1.5708$.
For the ideal cross ($\alpha \to 0$): $\lambda_{\max} = \frac{1}{2} L^2$, $n_c = 2$.

The eigenvalue method ignores closed loops in the network, which systematically underestimates the absolute threshold values compared with Monte Carlo (MC)~\cite{Grazioli2026}; therefore, for a correct comparison of trends, the percolation thresholds obtained by theoretical estimates and by computer simulation are normalized~\cite{Grazioli2026}:
\[\tilde{Y} = \frac{n_c - n_{\text{iso}}}{n_{\text{pc}} - n_{\text{iso}}}, \quad Y = \tilde{Y} \frac{n_{\text{pc}}}{n_c}.\]
Here and below, iso refers to the isotropic distribution, and pc refers to the ideal cross. The normalized curves for different distributions and thicknesses collapse onto a single universal dependence, which confirms that the analytical estimate correctly reproduces the trend.

The excluded areas of zero-thickness sticks (of length $l$) and discorectangles (with straight-section length $l$ and width $w$) differ by a constant $C = 4lw + \pi w^2$. This leads to a simple and elegant relation between the corresponding connectivity matrices $\mathrm{M}^{\text{stick}}$ and $\mathrm{M}^{\text{DR}}$.

The elements of the matrix for zero-thickness sticks have the form:
\[M^{\text{stick}}_{jk} = \sqrt{p_j p_k} \, l^2 |\sin(\theta_j - \theta_k)|.\]
The elements of the matrix for discorectangles are:
\[M^{\text{DR}}_{jk} = \sqrt{p_j p_k} \left( C + l^2 |\sin(\theta_j - \theta_k)| \right) = M^{\text{stick}}_{jk} + C \sqrt{p_j p_k}.\]
Introducing the vector $\mathbf{v} = (\sqrt{p_1}, \sqrt{p_2}, \dots, \sqrt{p_{N_\theta}})^T$, this relation can be written in matrix form:
\[M^{\text{DR}} = M^{\text{stick}} + C \mathbf{v} \mathbf{v}^T,\]
with $\operatorname{rank}(C\mathbf{v}{{\mathbf{v}}^{T}})=1$. The only nonzero eigenvalue of the matrix $C \mathbf{v} \mathbf{v}^T$ is equal to $C$, since the norm of the vector $\mathbf{v}$ is equal to one ($\|\mathbf{v}\|^2 = \sum p_i = 1$). The relation between the maximum eigenvalues of the matrices $\mathrm{M}^{\text{DR}}$ and $\mathrm{M}^{\text{stick}}$ has the form:
\[\lambda_{\max}^{\text{DR}} \approx \lambda_{\max}^{\text{stick}} + C.\]
The equality is approximate rather than exact because the vector $\mathbf{v}$ (determined by the probability distribution) does not coincide exactly with the principal eigenvector of the matrix $M^{\text{stick}}$.

The strict relation is given by the characteristic equation (a corollary of the matrix determinant lemma):
\[1 - C \sum_{k=1}^{N_\theta} \frac{v_k^2}{\lambda_{\max}^{\text{DR}} - \lambda_k^{\text{stick}}} = 0,\]
where $\lambda_k^{\text{stick}}$ are all eigenvalues of the stick matrix, and $v_k^2 = p_k$. The contribution of the remaining eigenvalues (the sum over $k$) gives a small correction (of second order of smallness) to the simple shift.

Since, within the approximation used by the authors of~\cite{Grazioli2026}, the percolation threshold is determined as $n_c = 1/\lambda_{\max}$, the approximate relation for the thresholds takes the form:
\[\frac{1}{n_c^{\text{DR}}} \approx \frac{1}{n_c^{\text{stick}}} + C,\]
or
\[n_c^{\text{DR}} \approx \frac{n_c^{\text{stick}}}{1 + C n_c^{\text{stick}}}.\]
The constant part of the excluded area $C$ linearly shifts the inverse threshold, weakening the relative contribution of the angular dependence. A natural question arises: how does the presence of this constant (finite thickness) affect the form of the normalized curves $\tilde{Y}$ and $Y$ that were used in~\cite{Grazioli2026} to compare trends?

Consider the limiting cases:
\begin{itemize}
    \item Case of disks ($\varepsilon = 1$, $l=0$): The stick matrix vanishes ($M^{\text{stick}} = 0$). The characteristic equation gives the strict equality $\lambda_{\max}^{\text{DR}} = C$, since all $\lambda_k^{\text{stick}} = 0$, and the mismatch of eigenvectors plays no role. Percolation threshold: $n_c = 1/C = 1/(\pi w^2)$.
\item  Case of thin sticks ($w \to 0$, $\varepsilon \to \infty$): The constant $C = 0$, and $M^{\text{DR}}$ exactly coincides with $M^{\text{stick}}$. The percolation threshold passes to the classical result for zero-thickness sticks.
\end{itemize}
Thus, although the relation between the matrices is algebraically simple (a rank-1 matrix), the relation between their principal eigenvalues (and percolation thresholds) is approximate because of the difference in the structure of the eigenvectors, except in the trivial limiting cases.

Substitute the relation for the thresholds into the formula for the first normalization from~\cite{Grazioli2026}:
\[\tilde{Y}^{\text{DR}}   = \tilde{Y}^{\text{stick}}   \frac{1 + C n^{\text{stick}}_{\text{pc}}}{1 + C n^{\text{stick}}_c}.\]
Since $n_c \leqslant n_{\text{pc}}$, the factor $\frac{1 + C n_{\text{pc}}}{1 + C n_c} \geqslant 1$. This means that in the presence of finite thickness (i.e., for $C > 0$) the curve $\tilde{Y}^{\text{DR}}$ will lie above the curve for sticks $\tilde{Y}^{\text{stick}}$.

Now apply the second normalization from~\cite{Grazioli2026} and take into account that in the eigenvalue method the percolation threshold is determined as $n_c = 1/\lambda_c$, where $\lambda_c \equiv \lambda_{\max}$ is the maximum eigenvalue of the connectivity matrix; the inverse threshold $1/n_c$ is exactly equal to $\lambda_c$:
\[Y = \frac{n_c n_{\text{pc}} - n_{\text{iso}} n_{\text{pc}}}{n_c n_{\text{pc}} - n_c n_{\text{iso}}}=
    \frac{1/n_{\text{iso}} - 1/n_c}{1/n_{\text{iso}} - 1/n_{\text{pc}}} = \frac{\lambda_{\text{iso}} - \lambda_c}{\lambda_{\text{iso}} - \lambda_{\text{pc}}}.\]
Thus, the second normalization is a linear normalization of the maximum eigenvalue between its limits: the isotropic distribution and the ideal cross. Such a representation makes obvious the universality of the $Y$ curves: any additive constants in the connectivity matrix (for example, $C = 4lw + \pi w^2$ for discorectangles, since $\lambda_c^{\text{DR}} \approx \lambda_c^{\text{stick}} + C$) cancel completely in the numerator and denominator, leaving only the purely orientational dependence.

From the obtained result follows a fundamental conclusion: the second normalization $Y$ completely compensates for the influence of finite thickness. Finite thickness changes the absolute values of the thresholds and distorts the first normalization $\tilde{Y}$, but the second normalization $Y$ “absorbs” this distortion, leaving only information about orientational ordering, cleaned of the geometric factor of thickness.

This conclusion is strictly valid within the approximation $M^{\text{DR}} \approx M^{\text{stick}} + C \mathbf{v} \mathbf{v}^T$. When using the exact eigenvalue method (without this approximation), the equality $Y^{\text{DR}} = Y^{\text{stick}}$ will hold to within fractions of a percent because of the imperfect coincidence of eigenvectors, which does not affect the principal conclusion about the universality of the second normalization.

\begin{figure}[!htb]
  \centering
  \includegraphics[width=0.6\textwidth]{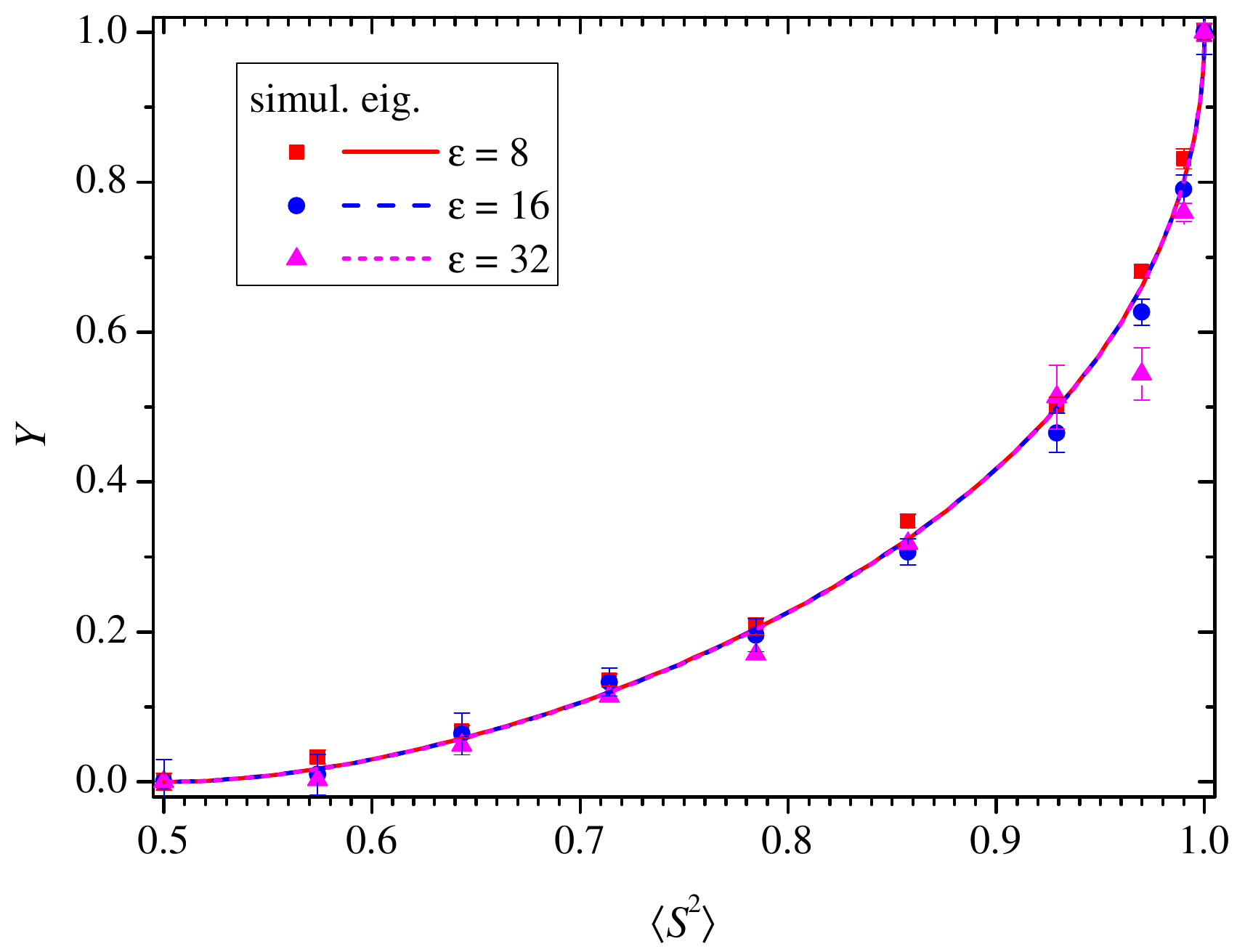}
  \caption{Dependence of the normalized percolation threshold on the second moment of the angular distribution. Lines are theoretical estimates; markers are computer simulation results}\label{fig:results}
\end{figure}

To verify universality, a computer simulation was performed. Discorectangles of unit length with aspect ratios 8, 16, and 32 were used; they were placed sequentially, independently, and randomly in a domain $L_D \times L_D$ with periodic boundary conditions. The domain sizes were $L_D = 8, 16, 24, 32$. The orientations of the discorectangles were specified by distribution~\eqref{eq:APDF-MC}. The union--find algorithm~\cite{Mertens2012} was used to determine the percolation threshold. $10^3$ independent realizations were used for system sizes 24 and 32, while $10^4$ realizations for sizes 8 and 16. Since the percolation probability was determined for a fixed number of particles, a convolution with the Poisson distribution was performed. Various criteria were used: the appearance of a wrapping cluster horizontally, vertically, and in both directions. A scaling analysis was performed. For the final estimate, the average values of the percolation thresholds obtained using the three criteria were used. The simulation results together with the theoretical estimate presented in Fig.~\ref{fig:results} confirm the presence of universality.

\bibliographystyle{plain}
\bibliography{percolation-cross}
\end{document}